# Enhancing Speech Quality through the Integration of BGRU and Transformer Architectures

**Eng. Souliman Alghnam[1] , Prof. Dr. Mohammad Alhussien[2] , Dr. Khaled Shaheen[3]**

[1]Postgraduate Student, Department of Electronics and Communications Engineering, Faculty of Mechanical and Electrical Engineering, Damascus University, E-Mail: soulimanalghnam@gmail.com

[2]Professor, Department of Electronics and Communications Engineering, Faculty of Mechanical and Electrical Engineering, Damascus University.

[3]Professor, Department of Electronics and Communications Engineering, Faculty of Mechanical and Electrical Engineering, Damascus University.

**Abstract**

Speech enhancement plays an essential role in improving the quality of speech signals in noisy environments. This paper investigates the efficacy of integrating Bidirectional Gated Recurrent Units (BGRU) and Transformer models for speech enhancement tasks. Through a comprehensive experimental evaluation, our study demonstrates the superiority of this hybrid architecture over traditional methods and standalone models. The combined BGRU-Transformer framework excels in capturing temporal dependencies and learning complex signal patterns, leading to enhanced noise reduction and improved speech quality. Results show significant performance gains compared to existing approaches, highlighting the potential of this integrated model in real-world applications. The seamless integration of BGRU and Transformer architectures not only enhances system robustness but also opens the road for advanced speech processing techniques. This research contributes to the ongoing efforts in speech enhancement technology and sets a solid foundation for future investigations into optimizing model architectures, exploring many application scenarios, and advancing the field of speech processing in noisy environments.

## 1. Introduction

Speech enhancement is a task to enhance the quality of speech signal in a noisy environment. Speech enhancement can be divided into two types: single channel and multichannel. Single channel is considered more challenging since only one microphone records the speech so the data observing is less than multichannel case. Speech enhancement is very important for many applications like automatic speech recognition (ASR), internet of things (IoT) which need interaction between human and machine by voice, and Hearing aids. Speech enhancement is a special case of speech separation problem, where the objective is to remove the background noise and predict the desired speech signal, speech enhancement can be mathematically expressed by the following equation:

$$y(t) = x(t) * h(t) + n(t)$$

Where $x(t)$ is the clean signal, $h(t)$ is the impulse response of the recording environment which acts as a filter, * is the convolution multiplication, $n(t)$ is the background noise. The aim of speech enhancement is to extract $x(t)$ from $y(t)$ without any information about $h(t)$ and $n(t)$.

## 2. Related Work

The first methods for solving the problem depended on signal processing techniques. Spectral subtracting (Verteletskaya & Simak, 2011) was the easy way to remove the noise from recorded speech signals by estimating the noise in the noisy signals using voice activity algorithms then calculate the noise spectrum then subtract it from the spectrum of the noisy signal. Weiner filter was used (Gomez & Kawahara, 2010; Krawczyk-Becker & Gerkmann., 2015) to estimate the noise in each frame then updating the filter parameters due to estimated noise level. CEEMDAN algorithm was used in (Melhem, et al., 2024) to denoise speech signals which is a new version of EMD algorithm. In (So, et al., 2017) authors could separate colored noise from speech using adaptive filtering methods, the most important of those filters is the Kalman Filter, which performs an optimal estimation of the system's statistical parameters, but at the expense of the computational complexity and the system's hardware. Acoustic feedback problem in hearing aids was addressed in (Lala, 2017) by integrating Least Mean Square (LMS) algorithm with adaptive filtering techniques. (Tarrab, 2020) offered a comparative study between statistical methods and auditory models for acoustic feedback suppression in binaural hearing aids, the results showed the superiority of statistical methods. All previous methods suffered from artificial noise and disability to some colored noises. Deep learning models offered a great improvement in speech enhancement, in (Abdulatif, et al., 2024), authors introduces a framework CMGAN for speech enhancement, that composed of a CNN and transformer to extract a good representation of speech, and a GAN to optimize the metric in non-differentiable way. The model is fed by time-frequency domain samples, using the real and imaginary parts of spectrogram. Testing the model on VOICE Bank dataset gave 3.41 for PESQ and 11.10 for SNR.

Authors in (Fan, et al., 2023) proposed a CompNet which consists of two modules, the first is a pre-processing time-domain network which is a Temporal Convolutional Neural Network (TCNN), and the second is T-F domain post-processing network which is a U-Net, then filtering the magnitude and refine the phase. CompNet is a multi-domain framework for speech enhancement, it enhances the waveform by the first network, then estimate the spectrum of the speech signal. The experiments were done on WSJ0-SI84 and VoiceBank + Demand datasets and the results were competitive.

MP-SENet is proposed in (Lu, et al., 2023), a TF-domain monaural SE model with parallel magnitude and phase spectra enhancing, the overall structure is composed of Encoder, Transformer and decoder the input of the network is spectrum and the decoder is consist of two parallel decoder one for magnitude and the other for phase. the results show the quality of enhanced speech is 3.5 for PESQ metric.

Authors in (Tai, et al., 2024) address the condition collapse problem in diffusion models to enhance the intelligibility of speech signal, so they propose a Diffusion-dropout Speech Enhancement method (DOSE), the method utilize two techniques, the first randomly dropping out samples during training to force the model to focus on condition factor, the second generating samples from adaptive prior derived from conditional factor instead of Gaussian distribution. The results were high quality.

in (Yang, et al., 2024), authors try to close the gap between simulated noise and real-world noise by building a new real-recorded and annotated microphone array speech and noise dataset. Benchmarking for speech enhancement and source localization algorithms in real scenarios were done, and the performance could reflect the capability of algorithms in real-world applications.

in (Zhang, et al., 2023) authors have devised an unconstrained speech enhancement and separation model (USES) that can handle denoising and dereverberation in diverse input conditions altogether, including variable microphone channels, sampling frequencies, signal lengths, and different environments. The results showed a high quality enhancement 3.46 for PESQ and 0.98 for STOI.

In this paper, we introduce a new model for speech enhancement based on two component BGRU and Transformer.

### 3.Proposed model

The model used depends essentially on two main components BGRU and Transformer.

**3.1.Bidirectional Gated Recurrent Units (BGRU)**: They are a type of neural network architecture designed for sequential data processing, such as in natural language processing and time series analysis. BGRU combine the advantages of Bidirectional RNNs with the gating mechanisms of GRUs, allowing them to capture dependencies in both forward and backward directions while effectively managing information flow through reset and update gates. This bidirectional nature enables them to better understand context and dependencies within a sequence, making them particularly useful for tasks requiring a comprehensive understanding of input data. Many studies improved the performance of speech separation models using BGRU like (Melhem, et al., 2023).

**3.2.Transformer:** The Transformer architecture revolutionized natural language processing by introducing the self-attention mechanism, enabling parallel processing of input sequences. Unlike RNNs and LSTMs, Transformers can capture long-range dependencies efficiently, making them ideal for tasks like speech enhancement. In speech enhancement, Transformers excel at denoising audio signals by learning complex patterns and relationships within the input spectrogram data. Their attention mechanisms focus on relevant parts of the input, allowing them to isolate noise from speech signals effectively. Transformers also facilitate easier parallelization and scalability compared to recurrent models, making them efficient for processing large amounts of audio data in real-time applications. The figure 1 shows the architecture of transformer where it is obvious that it totally depends on Attention mechanism.

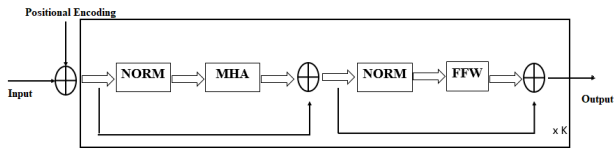

Figure 1 Transformer Architecture

### 3.3. Blockformer

It is a block of transformers, which uses the dual-path processing, it consists of two transformers, Intra-transformer to model the dependencies between samples in each frame, and inter-transformer which model the dependencies between each frame and others. The architecture of intra-transformer and inter-transformer is the same, but the permutation operation between them makes each one acts differently. The blockFormer is repeated R times.

In this model, as described in Figure 2, the input comprises the magnitude of the short-time Fourier transform, which is calculated using Hanning window of length 512 and hop size of 128 samples. Then it is processed by a Bidirectional Gated Recurrent Unit (BGRU) to generate a novel representation of the spectrogram. The output from the BGRU is subsequently forwarded to a Blockformer to compute a mask (m) for noise removal from the speech signal. By applying this mask to the magnitude of the speech spectrogram and utilizing the phase information from the mixture, the model reconstructs the time-domain signal. The loss function employed in this study is the Signal-to-Noise Ratio (SNR).

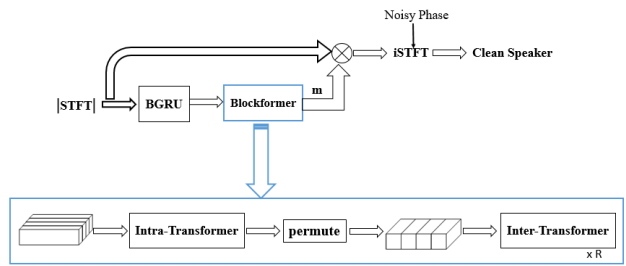

Figure 2. The proposed model for speech enhancement

### 4. Dataset

TIMIT corpus (Garofolo, et al., 1993) is widely known in the field of speech research, offers a rich dataset with recordings from diverse English dialects, aiding in speech recognition and enhancement studies. Some researchers (Melhem, et al., 2023; Melhem, et al., 2024) created a new version of TIMIT to use it in speech separation. It consists of recordings of 630 speakers of eight major dialects of American English, each reading ten phonetically rich sentences. TIMIT is valuable for its phonetically transcribed annotations and its high quality, making it a crucial resource for training and testing speech processing systems. Researchers utilize TIMIT to evaluate algorithms, study acoustic-phonetic characteristics, and advance speech technology. Its structured design and detailed annotations have made TIMIT a cornerstone in the study of speech processing and a benchmark dataset in the field. We also utilize the Non-Speech Noise Dataset (HuCorpus, n.d.) to produce a noisy file comprising 100 varieties of noise such as wind, babble, engines, street, and more. These realistic noise types prove highly beneficial for training our model effectively. We leverage 6300 speech files sourced from TIMIT along with 100 noise files to generate 630k noisy files with different SNR values from -10 dB to 10 dB. These files are then divided into three sets: 70% for training, 20% for validation, and 10% for testing purposes.

The assessment criteria employed in this research encompass Signal-to-Noise Ratio (SNR), Perceptual Evaluation of Speech Quality (PESQ), and Short Time Objective Intelligibility (STOI).

## 5. Results and Discussion

To evaluate our model's performance, we test it across various noise types spanning SNR values between -10 and 10. The Figure 3 displays five curves illustrating the relationship between the output's SNR and the input's SNR.

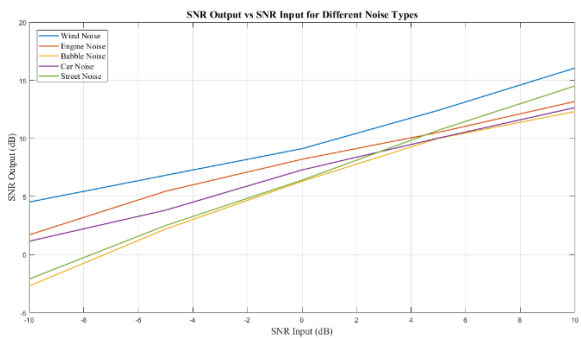

Figure 3 the performance of our model in different types of noise

Among the noise types evaluated, the curve for Babble noise exhibited the least enhancement, whereas the curve for Wind noise demonstrated the most significant improvement. This discrepancy can be attributed to the presence of speech within Babble noise, posing challenges for noise removal. In contrast, Wind noise, characterized as stationary noise, was easier to effectively eliminate, thus yielding superior results.

Figure 4 depicts a sample of a noisy signal with SNR of -5 dB, contrasted with an enhanced version boasting an SNR of 3.8 dB. Additionally, Figure 5 presents the spectrum of both the noisy and enhanced signals.

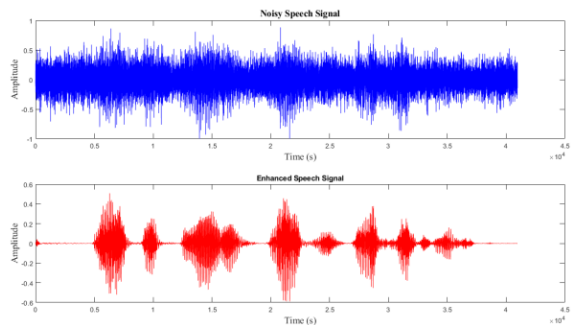

Figure 4. sample of noisy and enhanced speech signal

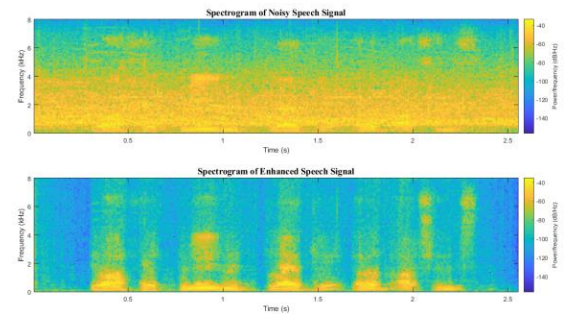

Figure 5. the spectrum of noisy and enhanced speech signals

To investigate the fluctuation in our model's performance under varying signal-to-noise ratios (SNRs), we employ box plots for Short Time Objective Intelligibility (STOI) and Mean Opinion Score (MOS). The SNR range considered spans from -10 dB to 10 dB. These graphical representations are detailed in Figure 6,7).

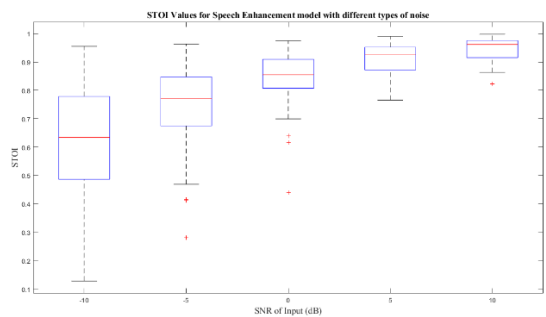

Figure 6. box plot for STOI for different values of SNR

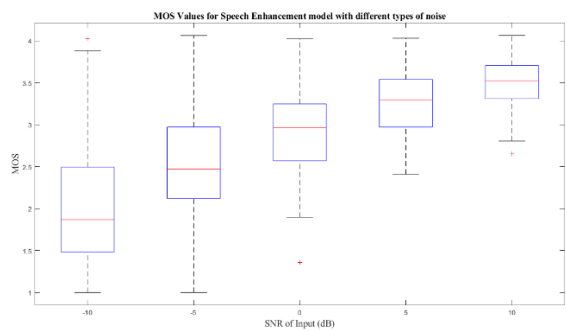

Figure 7. box plot for MOS versus different values of SNR.

The preceding pair of box plots illustrate the effectiveness of our model. As depicted in Figure 6 and Figure 7, the performance variance of our model is notably narrow at high input SNR levels, with median values around 0.95 for STOI and 3.5 for MOS. Conversely, the variance widens at low SNR (-10 dB), which is expected due to the higher energy of noise compared to the speech signal in such conditions. Nevertheless, even under these circumstances, the median values remain satisfactory, approximately 1.9 for MOS and 0.62 for STOI.

Three state-of-the-art speech enhancement models, namely CMGAN (Abdulatif, et al., 2024), CompNet (Fan, et al., 2023) and DOSE (Tai, et al., 2024), were trained on the identical dataset for comparative analysis with our model. The outcomes of this comparison are presented in the Table 1.

Table 1. Comparing our model with three state of the art speech enhancement models.

|  | SNR | PESQ | STOI |
|---|---|---|---|
| CMGAN | 3.56 | 3.21 | 0.71 |
| DOSE | 3.71 | 3.14 | 0.59 |
| CompNet | 3.01 | 2.91 | 0.65 |
| Our model | **3.83** | **3.64** | **0.78** |

The superior performance of our model over three others underscores the successful integration of BGRU and Transformer in representing speech signals effectively, highlighting its competitive edge in signal processing.

**6. Conclusion**

In conclusion, the utilization of Bidirectional Gated Recurrent Units (BGRU) and Transformer models in speech enhancement has demonstrated promising results. Through effective signal representation and processing, these models have shown significant improvements in noise reduction and speech quality. The great integration of BGRU and Transformer architectures has enhanced the robustness and performance of speech enhancement systems, introducing the way for advanced applications in noisy environments. Future research could focus on optimizing the architecture further, exploring different hyper parameters, and investigating real-world deployment scenarios to fully leverage the potential of these models in enhancing speech quality across various domains.